\documentclass[final]{elsart5p}
\usepackage{amsmath}
\usepackage{amssymb}
\usepackage{graphicx}
\usepackage{subfigure}
\usepackage{latexsym}
\usepackage[square,numbers,sort&compress,comma]{natbib}
\usepackage{mathptm} 
\usepackage{psfrag}
\usepackage{amssymb}

\begin{document}

\begin{frontmatter}

 \title{
        Modified Associate Formalism without Entropy Paradox:\\
        Part I. Model Description
       }

 \author{
        Dmitry N. Saulov\corauthref{cor}\thanksref{label1}
        }
        \ead{d.saulov@uq.edu.au},\quad
 \author{
        Igor G. Vladimirov\thanksref{label1}
        },\quad
 \author{
        A. Y. Klimenko\thanksref{label1}
        }

 \corauth[cor]{
               Corresponding author:
               School of Engineering,
               The University of Queensland,
               St. Lucia, QLD 4072, Australia.\newline
               Phone: (07)3365-3677
               Fax:   (07)3365-3670
              }

\address[label1]{School of Engineering, The University of Queensland, AUSTRALIA}

\journal {Journal of Alloys and Compounds} \volume {473(1–2)} \pubyear{2009} \firstpage{167} \lastpage{175 }

\begin{abstract}
A Modified Associate Formalism is proposed for thermodynamic modelling of
solution phases. The approach is free from the entropy paradox described by
L\"{u}ck et al. (Z. Metallkd. 80 (1989) pp. 270--275). The model is considered
in its general form for an arbitrary number of solution components and an
arbitrary size of associates. Asymptotic behaviour of chemical activities of
solution components in binary dilute solutions is also investigated.
\end{abstract}



\end{frontmatter}

\section{Introduction}
The associate model in its various modifications has been successfully used for modelling
solution phases of metallurgical and chemical engineering
interest~\cite{Sommer82_2,Hastie85,Besmann02,Besmann06,Yazhenskikh_06,Yazhenskikh_06_1,Yazhenskikh_07}.
Besmann and Spear \cite{Besmann02}, for example, utilised the modified associated species
model for glasses used in nuclear waste disposal. Recently, Yazhenskikh \emph{et al.}
\cite{Yazhenskikh_06,Yazhenskikh_06_1,Yazhenskikh_07} have successfully applied the
associate species model to model melting behaviour of coal ashes which is an important
problem in coal gasification technologies. Good agreement between model predictions and
available experimental data was reported.

According to the classical associate model described by Prigogine and Defay
\cite{PrigogineDefay}, the strong interactions result in the formation of stable
configurations of mixing particles, the so-called association complexes or, briefly,
associates. Those particles that are not involved in the formation of associates are
called free particles or, interchangeably, monoparticles. The associated solution is then
considered to be an ideal solution of monoparticles and different associates. For
example, a binary associated solution of components $A$ and $B$, in which only
$AB$-associates are formed, is considered to be a ternary ideal solution of the
$A$-monoparticles, $B$-monoparticles and $AB$-associates.

The Gibbs free energy $G$ of the associated solution of $n_{A}$ moles of the solution
component $A$ and $n_{B}$ moles of $B$ is then given by
\begin{equation}
\label{eq associated G}
    G
    =
    n_{A_1} g_{A_1}
    +
    n_{B_1} g_{B_1}
    +
    n_{AB}  g_{AB}
    -
    TS_\text{conf},
\end{equation}
where the configurational entropy of mixing $S_\text{conf}$ is expressed as
\begin{equation}
\label{eq associated S_config}
    S_\text{conf}
    =
    -R
    \left(
        n_{A_1}
        \ln x_{A_1}
        +
        n_{B_1}
        \ln x_{B_1}
        +
        n_{AB}
        \ln
        x_{AB}
    \right).
\end{equation}
Here, $n_{A_1}$, $n_{B_1}$, $n_{AB}$ are the mole numbers and $g_{A_1}$,
$g_{B_1}$, $g_{AB}$ are the molar Gibbs free energies of the $A$-monoparticles,
$B$-monoparticles and $AB$-associates, respectively; $T$ is the absolute
temperature and $R$ is the universal gas constant. The molar fractions
$x_{A_1}$, $x_{B_1}$ and $x_{AB}$ are defined in a usual way. For example,
$x_{A_1} = n_{A_1} / (n_{A_1} + n_{B_1}+n_{AB})$. The other molar fractions are
defined similarly.

The equilibrium values of the mole numbers $n_{A_1}$, $n_{B_1}$ and $n_{AB}$
are determined by minimising the Gibbs free energy $G$ given by Eq.~(\ref{eq
associated G}), subject to the mass balance constraints
\begin{equation}
\label{}
\begin{array}{l}
    n_{A} = n_{A_1} + n_{AB},\\
    n_{B} = n_{B_1}+n_{AB}.
\end{array}
\end{equation}
The adjustable parameter of the model is the molar Gibbs free energy $\Delta
g_{AB}$ of the reaction
\begin{equation}
\label{}
    A + B
    \rightleftarrows
    AB.
\end{equation}

Besmann and Spear \cite{Besmann02} considered an ideal mixture of
\emph{associate species} (instead of monoparticles and associates). The
stoichiometry of the associate species was specified so that all the species
contain two non-oxygen atoms per formula unit.  In this approach, contributions
of different species to the configurational entropy of mixing are equally
weighted.

L\"{u}ck \emph{et al.} \cite{Luck_entropy_paradox} described a remarkable
feature of the configurational entropy of mixing given by Eq.~(\ref{eq
associated S_config}). This feature was referred to as an entropy paradox and
is briefly described in the next section.

\section{Entropy paradox\label{sec paradox}}
\begin{figure}
\begin{center}
\includegraphics[width=\columnwidth]{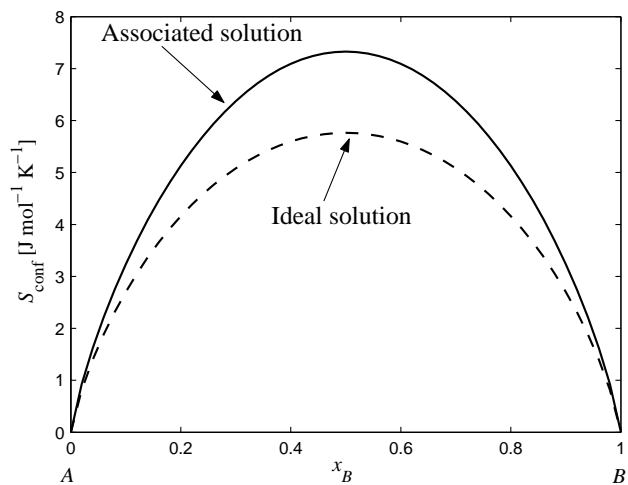}%
\caption {
    The configurational entropy of mixing for the associated solution, in which only
    the ${AB}$-associates are formed, in the high temperature limit compared with the ideal
    entropy of mixing.
\label{fig entropy ideal vs associated}}%
\end{center}
\end{figure}

L\"{u}ck \emph{et al.} considered the high temperature limit for the associated solution.
The temperature $T$ is assumed to be so high that the entropy term plays a dominant role
in the Gibbs free energy of the solution and the enthalpy changes on forming different
associates can be neglected. The configurational entropy of mixing given by Eq.~(\ref{eq
associated S_config}) in the high temperature limit is higher than that used in the
regular solution model (the entropy of ideal mixing of the solution components). For
example, the configurational entropy of mixing of the associated solution in which only
the ${AB}$-associates are formed is compared with the ideal entropy of mixing in Fig.
\ref{fig entropy ideal vs associated}.
At the same time, the formation of associates models short range
ordering in the solution. As pointed out by L\"{u}ck \emph{et
al.}~\cite{Luck_entropy_paradox}, it is a paradoxical result that
the configurational entropy, which is a measure of disorder, appears
to be higher in a solution with ordering than in a completely
disordered solution.

Another interpretation of this entropy paradox was described by Pelton \emph{et
al.} \cite{Pelton_00}. Consider the binary associated solution where only
${AB}$-associates are formed and assume that there are no Gibbs free energy
changes on forming ${AB}$-associates from monoparticles, so that $\Delta
g_{AB}=0$. In this case, the configurational entropy of mixing of the solution
should be equal to that of an ideal solution, since no interactions between
mixing particles are assumed.  Eq.~(\ref{eq associated S_config}) however,
leads to higher values for the configurational entropy of mixing that reduces
to the ideal configurational entropy of mixing only when $\Delta
g_{AB}=+\infty$; see also Fig.~\ref{fig entropy ideal vs associated}. The
overestimation of the configurational entropy can result in either
underestimation of the non-configurational entropy or overestimation of the
enthalpy of mixing or both. This, in tern, can undermine the predictive
capabilities of the model.

As pointed out by L\"{u}ck \emph{et al.}~\cite{Luck_entropy_paradox} and later
by Pelton \emph{et al.}~\cite{Pelton_00}, the expression for the
configurational entropy of mixing used in the quasichemical model does reduce
to the ideal configurational entropy of mixing when $\Delta g_{AB}=0$. In the
next section, we propose the Modified Associate Formalism which is free of the
entropy paradox. The paradox is resolved by distinguishing between all possible
spatial arrangements of particles in an associate that have not been taken into
account in previous associate models.

\section{Model assumptions\label{sec assumptions}}
The model proposed in this paper is based on the following assumptions.
\begin{enumerate}
\item[1)]
Similarly to the classical associate model \cite{PrigogineDefay}, we assume
that interactions between mixing particles result in the formation of
associates which are in a stable dynamic equilibrium with each other. The
associates of the model are understood as a tool for modelling short-range
interactions between mixing particles. The associates of the model, however,
may represent real associated complexes present in solution phases.
\item[2)]
We assume that the associates do not interact with each other and are uniformly
distributed (ideally mixed) over a lattice. Equivalently, the occupancies of
the sites of the associate lattice are stochastically independent and have
identical probability distributions.
\item[3)]
In contrast to the classical model, all pure solution components and the
chemical solution of these components are treated in a unified way. More
precisely, we assume that the solution and all its components consist of
noninteracting associates of the same size, so that the associates are composed
of the same number of particles. In this approach, the contributions of
different associates to the configurational entropy of mixing are equally
weighted.
\item[4)]
Following the convention (see, for example, Ref. \cite{Landau_stat_phys1} for
more details), we also assume  that particles of the same type are
indistinguishable, while particles of different types and particle sites within
an associate are distinguishable.
\end{enumerate}

\section{Model description\label{sec description}}
For simplicity of exposition,  we exemplify the model by considering a binary solution
$A-B$ where the associates are composed of \emph{three} particles. Let $n_{A}$ and
$n_{B}$ be the mole numbers of $A$ and $B$ particles in the solution. Since, by
assumption 4) above, the  particle sites within an associate are distinguishable, we also
assume that they are numbered. If the 1st and 2nd sites in an associate are both occupied
by $A$-particles, while the 3rd one is occupied by a $B$-particle, such an associate is
said to be of type $[AAB]$. Other types of associates are defined similarly. Thus, in the
solution considered, there are $2^3 = 8$ different types of associates, $    [AAA]$, $
    [BAA]$, $
    [ABA]$, $
    [AAB]$, $
    [BBA]$, $
    [BAB]$, $
    [ABB]$, $
    [BBB]
$. It is important to note that \emph{all} these types should be taken into
account in calculating the configurational entropy.

Let $n_{[ijk]}$ be the mole number of $[ijk]$-associates, where the triplet of
symbolic indices $i,j,k = A,B$ specify the associate type. Then the mass
balance constraints take the form
\begin{equation}
\label{eq mass balance}
\begin{array}{lcl}
    n_{A}
    & = &
    3n_{[AAA]}\\
& + &
    2(
        n_{[BAA]}
        +
        n_{[ABA]}
        +
        n_{[AAB]}
    )\\
& + &
    n_{[BBA]}
    +
    n_{[BAB]}
    +
    n_{[ABB]},\\
n_{B}
    & = &
    3n_{[BBB]}\\
& + &
    2(
        n_{[ABB]}
        +
        n_{[BAB]}
        +
        n_{[BBA]}
    )\\
&+ &
    n_{[AAB]}
    +
    n_{[ABA]}
    +
    n_{[BAA]}.
\end{array}
\end{equation}
By a standard combinatorial argument, the number of available microstates $\Omega$ is
\begin{equation}
    \label{eq omega}
    \Omega
    =
    \frac
    {
        \left(
            N_{\circ}
            \sum\limits_{i,j,k={A,B}}
            n_{[ijk]}
        \right)!
    }
    {
        \prod\limits_{i,j,k={A,B}}
        \left(
            N_{\circ}\,
            n_{[ijk]}
        \right)!
    },
\end{equation}
where $N_{\circ}$ is the Avogadro number. Under assumption 2) of ideal mixing
of associates, the configurational entropy of mixing $S_\text{conf}$ is
\begin{equation}
\label{eq S config 0}
    S_\text{conf}
    =
    k_\text{B}\ln \Omega~,
\end{equation}
where $k_\text{B}$ is Boltzmann's constant. Applying the Stirling
formula to Eqs.~(\ref{eq omega}) and (\ref{eq S config 0}),
\begin{equation}\label{eq S config}
    S_\text{conf}
    =
    -R
    \sum\limits_{i,j,k={A,B}}
    n_{[ijk]}
    \ln x_{[ijk]},
\end{equation}
where
$$
    x_{[ijk]}
    =
    \frac
    {n_{[ijk]}}
    {
        \sum\limits_{i',j',k' = A,B}\
        n_{[i'j'k']}
    }
$$
is the molar fraction of the $[ijk]$-associates. Therefore,  the
Gibbs free energy of the solution is given by
\begin{equation}\label{eq G}
    G
    =
    \sum\limits_{i,j,k={A,B}}
    n_{[ijk]}
    \left(
        g_{[ijk]}
        +
        RT
        \ln x_{[ijk]}
    \right),
\end{equation}
where $g_{[ijk]}$ is the molar Gibbs free energy of $[ijk]$-associates. In one
mole of pure solution component $A$ there are $1/3$ moles of
$[AAA]$-associates. Hence, $g_{[AAA]}=3g_{A}$, where $g_{A}$ is the molar Gibbs
free energy of the pure solution component $A$. Similarly, $g_{[BBB]}=3g_{B}$,
where $g_{B}$ is the molar Gibbs free energy of the pure solution component
$B$.

The associates, which consist of both $A$ and $B$ particles, will be referred
to as \emph{mixed associates}. For the binary solution considered, there are
$8-2 = 6$ types of mixed associates. Their molar Gibbs free energies
$g_{[ijk]}$, with $[ijk]\ne [AAA], [BBB]$, are adjustable parameters of the
model. These parameters can depend on the temperature $T$ as
\begin{equation}
    g_{[ijk]}
    =
    h_{[ijk]}
    -
    Ts_{[ijk]},
\end{equation}
where $h_{[ijk]}$ and $s_{[ijk]}$ are the molar enthalpy and entropy of
${[ijk]}$-associates, respectively. Alternatively, as adjustable parameters of the model,
one can employ the Gibbs free energies of the reactions of forming mixed associates from
\emph{pure} $[AAA]$ and $[BBB]$-associates. For example, instead of $g_{[AAB]}$, the
Gibbs free energy $\Delta g_{[AAB]}$ of the following reaction can be used,
\begin{equation}
\label{eq reaction}
    \frac{2}{3}[AAA]
    +
    \frac{1}{3}[BBB]
    \rightleftarrows
    [AAB],
\end{equation}
so that
\begin{equation}
\label{eq Delta g}
    \Delta g_{[AAB]}
    =
    g_{[AAB]}
    -
    \frac{2}{3} g_{[AAA]}
    -
    \frac{1}{3} g_{[BBB]}.
\end{equation}
The other Gibbs free energies $\Delta g_{[ijk]}$ are defined
similarly.

Note that Eqs.~(\ref{eq mass balance}) define $n_{[AAA]}$ and $n_{[BBB]}$ as
linear functions of the mole numbers of the solution components $n_{A}$ and
$n_{B}$ and of the mole numbers $n_{[ijk]}$ of the mixed associate types with
$[ijk]\ne [AAA], [BBB]$. The mole numbers of the latter associates, which
consist of both $A$ and $B$ particles, are the internal variables of the model.
These are determined by minimising the Gibbs free energy of the solution at
constant $n_{A}$ and $n_{B}$, subject to the mass balance constrains of
Eqs.~(\ref{eq mass balance}). The equilibrium values of $n_{[ijk]}$, with
$[ijk]\ne {[AAA],[BBB]}$, are found from
\begin{equation}
\label{stationarity}
    \left(
        \frac
        {\partial G}
        {\partial n_{[ijk]}}
    \right)_{n_{A},\, n_{B},\, n_{[i'j'k']}}
    =0,
\end{equation}
where the derivative in $n_{[ijk]}$ is calculated for fixed $n_A$, $n_B$ and
fixed five variables $n_{[i'j'k']}$, with $[i'j'k']\neq [ijk], [AAA],[BBB]$. By
the chain rule, Eq.~(\ref{stationarity}) reads
\begin{equation}
\label{eq dG dn1}
    \frac
    {\partial G}
    {\partial n_{[AAA]}}
    \frac
    {\partial n_{[AAA]}}
    {\partial n_{[ijk]}}
    +
    \frac
    {\partial G}
    {\partial n_{[BBB]}}
    \frac
    {\partial n_{[BBB]}}
    {\partial n_{[ijk]}}
    +
    \frac
    {\partial G}
    {\partial n_{[ijk]}} = 0.
\end{equation}
Therefore, combining the last equation with Eqs.~(\ref{eq mass balance}) and (\ref{eq G})
gives
\begin{equation}
\label{eq equlibrium conditions}
    \frac
    {n_{[ijk]}}
    {
        (n_{[AAA]})^{\alpha_{[ijk]}/3}
        (n_{[BBB]})^{\beta_{[ijk]}/3}
    } =
    \exp
    \left(
        -\frac
        {\Delta g_{[AAB]}}
        {RT}
    \right).
\end{equation}
Here, $\alpha_{[ijk]}$ and $\beta_{[ijk]}$ stand for the numbers of
$A$ and $B$ particles in the $[ijk]$-associate, respectively. For
example, $\alpha_{[AAB]}=2$ and $\beta_{[AAB]}=1$.

Recall that Eqs.~(\ref{eq mass balance}) define $n_{[AAA]}$ as a
linear function of $n_A$ and $n_{[ijk]}$'s, and $n_{[BBB]}$ as a
linear function of $n_B$ and $n_{[ijk]}$'s, where $[ijk]\ne
[AAA],[BBB]$. The chemical potential $\mu_{A}$ of the solution
component $A$ is calculated as follows
\begin{equation}
\label{eq mu}
\begin{array}{lll}
    \mu_{A}
& = &
    \left(
        \frac
        {\partial G}
        {\partial n_{A}}
    \right)_{n_{B}}\\
& = &
    \sum\limits_{[ijk]}
    {
        \frac
        {\partial G}
        {\partial n_{[ijk]}}
        \frac
        {\partial n_{[ijk]}}
        {\partial n_{A}}
    }\\
& + &
    \frac
    {\partial G}
    {\partial n_{[AAA]}}
    \left(
        \frac
        {\partial n_{[AAA]}}
        {\partial n_{A}}
        +
        \sum\limits_{[ijk]}
        {
            \frac
            {\partial n_{[AAA]}}
            {\partial n_{[ijk]}}
            \frac
            {\partial n_{[ijk]}}
            {\partial n_{A}}
        }
    \right)\\
& + &
    \frac
    {\partial G}
    {\partial n_{[BBB]}}
    \left(
        \frac
        {\partial n_{[BBB]}}
        {\partial n_{A}}
        +
        \sum\limits_{[ijk]}
        {
            \frac
            {\partial n_{[BBB]}}
            {\partial n_{[ijk]}}
            \frac
            {\partial n_{[ijk]}}
            {\partial n_{A}}
        }
    \right)\\
& = &
    \frac
    {\partial G}
    {\partial n_{[AAA]}}
    \frac
    {\partial n_{[AAA]}}
    {\partial n_{A}}
    +
    \sum\limits_{[ijk]}
    \left(
        \frac
        {\partial G}
        {\partial n_{[AAA]}}
        \frac
        {\partial n_{[AAA]}}
        {\partial n_{[ijk]}}
    \right.\\
& &
    \left.
        +
        \frac
        {\partial G}
        {\partial n_{[BBB]}}
        \frac
        {\partial n_{[BBB]}}
        {\partial n_{[ijk]}}
        +
        \frac
        {\partial G}
        {\partial n_{[ijk]}}
    \right)
    \frac
    {\partial n_{[ijk]}}
    {\partial n_{A}}.
\end{array}
\end{equation}
Here, the sums are taken over $i,j,k={A,B}$ such that $[ijk]\neq
[AAA],[BBB]$. Using Eqs.~(\ref{eq G}) and (\ref{eq dG dn1}), one
verifies that
\begin{equation}
\label{eq mu1}
    \mu_{A}
 =
    \frac
    {\partial G}
    {\partial n_{[AAA]}}
    \frac
    {\partial n_{[AAA]}}
    {\partial n_{A}}
     =
    g_{A}
    +
    \frac{1}{3}
    RT
    \ln
    x_{[AAA]}.
\end{equation}
The chemical potential $\mu_{B}$ of the solution component $B$ is calculated in a similar
way.

Note that $[AAB]$, $[ABA]$ and $[BAA]$ associates differ in the
spatial arrangement of the constituent particles, though they have
the same ``chemical composition'' $A_2B$. The same distinction holds
for $[BBA]$, $[BAB]$ and $[ABB]$ associates of common composition
$AB_2$. Let us assume, for a moment, that only \emph{one} spatial
arrangement of associates is allowed for each composition, while the
other arrangements  are prohibited energetically, for example,
$g_{[BAA]} = g_{[ABA]} = +\infty$ and $g_{[BAB]} = g_{[ABB]} =
+\infty$. In this case, the proposed formalism reduces to the
associate species model \cite{Besmann02}, in which associate species
are composed of three particles.

Unlike previous modifications of the associate model, we take into account all possible
spatial arrangements of particles in an associate. If two or more associates of different
types are spatially symmetric to each other, then their molar Gibbs free energies are
considered equal. For example, 2-particle associates $[AB]$ and $[BA]$ are symmetric and
therefore are endowed with equal Gibbs energies $g_{[AB]}=g_{[BA]}$.

\begin{figure}
\begin{center}
    \subfigure
    [Particle sites are arranged along a (horizontal) line.
    The associates {$[BAA]$} and  {$[AAB]$} are mirror reflections of
    each other about axis $a$; they are not symmetric, however,  to {$[ABA]$}.]
    {\label{fig line}
    \includegraphics
    [width=6cm]
    {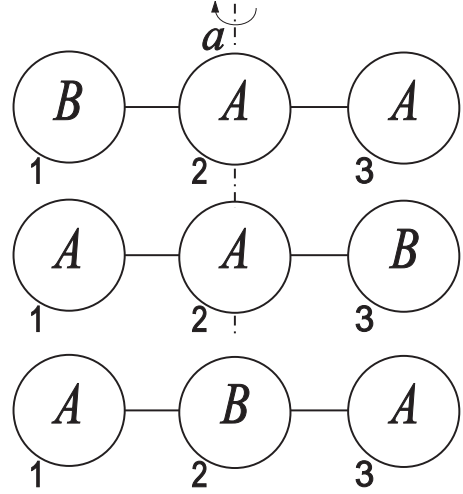}
} \hspace{5mm}
    \subfigure
    [Particle sites are arranged at the vertices of an equilateral triangle.
    All associates are symmetric to each other by rotation about axis $b$,
    perpendicular to the plane of the triangle.]
    {\label{fig triangle}
    \includegraphics
    [width=6cm]
    {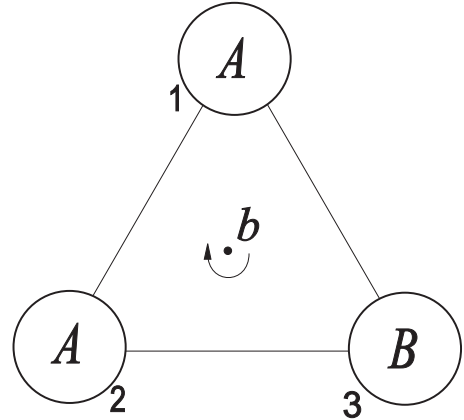}
} \caption {Schematic views of different arrangements of particle
sites. \label{f:combustion linking}}
\end{center}
\end{figure}

In more complex cases, however, Gibbs free energy levels  are
ascribed to associates depending on the spatial arrangement of
particles in them, so that both energy splits and multiple levels
may occur for associates of common chemical composition.
 For example, if particle sites in 3-particle associates
are arranged linearly as shown in Fig.~\ref{fig line}, then
$g_{[AAB]} = g_{[BAA]}$ while $g_{[ABA]}$ can be different.
Alternatively, if the particle sites are arranged as in
Fig.~\ref{fig triangle}, the three associate types are all symmetric
to each other, and hence, $g_{[AAB]} = g_{[ABA]} = g_{[BAA]}$.

Note that no assumptions on spatial arrangements of particle sites within an associate
have been made so far in the framework of the proposed model. In general, it is
impossible to determine in advance the number of energy levels and their multiplicities
for the associates of a particular composition. Furthermore, such associates may have the
same Gibbs free energy of formation, even if they are not spatially symmetric. As a
reasonable  initial approximation, all associates of a given composition can be endowed
with the same Gibbs energy of formation. This assumption can be refined subsequently in
the process of thermodynamic model optimisation for real chemical systems, if use of
several energy levels appears to provide a better fit to experimental data.

For the rest of this section and also in Sections~\ref{sec ideal case} and \ref{sec
dilute}, we assume, for simplicity, that the molar Gibbs free energy of an associate type
is completely specified by its chemical composition,
\begin{equation}
\label{eq equal delta g}
\begin{array}{l}
    \Delta g_{[AAB]}
    =
    \Delta g_{[ABA]}
    =
    \Delta g_{[BAA]}
    \equiv
    \Delta g_{2,1},\\
    \Delta g_{[BBA]}
    =
    \Delta g_{[BAB]}
    =
    \Delta g_{[ABB]}
    \equiv
    \Delta g_{1,2}.
\end{array}
\end{equation}
Here, the subscript ``2,1'' signifies that the associate consists of
two $A$-particles and one $B$-particle, with ``1,2'' and similar
indices understood appropriately. More general case is considered in
Section~\ref{sec general form}.

Using Eqs.~(\ref{eq equlibrium conditions}) and (\ref{eq equal delta
g}),
\begin{equation}\label{eq mole numbers}
\begin{array}{l}
    n_{[AAA]}
    \equiv
    n_{3,0},\\
    n_{[AAB]}
=
    n_{[ABA]}
=
    n_{[BAA]}
    \equiv
    n_{2,1}/3,\\
    n_{[BBA]}
    =
    n_{[BAB]}
    =
    n_{[ABB]}
    \equiv
    n_{1,2}/3,\\
    n_{[BBB]}
    \equiv
    n_{0,3}.
\end{array}
\end{equation}
In terms of molar fractions, Eqs.~(\ref{eq mole numbers}) read
\begin{equation}
\label{eq mole fractions}
\begin{array}{l}
    x_{[AAA]}
    \equiv
    x_{3,0},\\
    x_{[AAB]}
    =
    x_{[ABA]}
    =
    x_{[BAA]}
    \equiv
    x_{2,1}/3,\\
    x_{[BBA]}
    =
    x_{[BAB]}
    =
    x_{[ABB]}
    \equiv
    x_{1,2}/3,\\
    x_{[BBB]}
    \equiv
    x_{0,3}~.
\end{array}
\end{equation}
Substitution of Eqs.~(\ref{eq mole numbers}) and (\ref{eq mole fractions}) into
Eq.~(\ref{eq G}) gives
\begin{equation}\label{eq G1}
\begin{array}{lll}
    G
    &= &
    n_{3,0} g_{3,0}
    +
    n_{0,3} g_{0,3}
    +
    n_{2,1} g_{2,1}
    +
    n_{1,2} g_{1,2}\\
    &+&
    RT
    \left(
        n_{3,0}
        \ln
        x_{3,0}
        +
        n_{0,3}
        \ln
        x_{0,3}
    \right)\\
&+&
    RT
    \left(
        n_{2,1}
        \ln
        \left(
            \frac
            {x_{2,1}}
            {3}
        \right)
        +
        n_{1,2}
        \ln
        \left(
            \frac
            {x_{1,2}}
            {3}
        \right)
    \right).
\end{array}
\end{equation}
The mass balance constraints of Eqs.~(\ref{eq mass balance}) reduce
to
\begin{equation}\label{eq mass balance1}
\begin{array}{lll}
    n_{A}
    &=&
    3n_{3,0}
    +
    2n_{2,1}
    +
    n_{1,2},\\
    n_{B}
    &=&
    3n_{0,3}
    +
    2 n_{1,2}
    +
    n_{2,1}.
\end{array}
\end{equation}
The equilibrium values of the associate mole numbers are determined by
\begin{equation}\label{eq equilibrium conditions1}
\begin{array}{l}
    \frac
    {n_{2,1}}
    {
        n_{3,0}^{2/3}
        n_{0,3}^{1/3}
    }
    =
    3
    \exp
    \left(
        -\frac
        {\Delta g_{2,1}}
        {RT}
    \right),\\
    \frac
    {n_{1,2}}
    {
        n_{3,0}^{1/3}
        n_{0,3}^{2/3}
    }
    =
    3
    \exp
    \left(
        -\frac
        {\Delta g_{1,2}}
        {RT}
    \right).
\end{array}
\end{equation}
%

\section{Case $\Delta g_{2,1}=0$ and $\Delta g_{1,2}=0$ \label{sec ideal case}}
Recalling Eq. (\ref{eq equal delta g}), consider the situation where there are no Gibbs
free energy changes of forming different associates, that is, $\Delta g_{2,1}=0$ and
$\Delta g_{1,2}=0$. In this case, the $A$ and $B$ particle species mix ideally. The
composition of a randomly selected triplet of particles follows the binomial
distribution, well-known in probability theory; see, for example, Ref.~\cite{Feller_1}.
More precisely, the probabilities of choosing an associate of particular compositions,
or, equivalently, the molar fractions of appropriate associates, are
\begin{equation}\label{eq probabilities}
\begin{array}{ll}
    x_{3,0}
    &=
    x_{A}^3,\\
    x_{2,1}
    &=
    3x_{A}^2 x_{B},\\
    x_{1,2}
    &=
    3x_{A} x_{B}^2,\\
    x_{0,3}
    &=
    x_{B}^3.
\end{array}
\end{equation}
Here, $x_{A}$ and $x_{B}$ are the molar fractions of $A$ and $B$
particles. Since the total mole number of associates $n_\text{tot}$
is equal to $(n_{A} + n_{B})/3$, then the mole numbers of associates
are given by
\begin{equation}\label{eq mole numbers1}
\begin{array}{lll}
    n_{3,0}
    &=&
    \frac{n_{A}^3}{3(n_{A}+n_{B})^2},\\
    n_{2,1}
    &=&
    \frac{n_{A}^2 n_{B}}{(n_{A}+n_{B})^2},\\
    n_{1,2}
    &=&
    \frac{n_{A} n_{B}^2}{(n_{A}+n_{B})^2},\\
    n_{0,3}
    &=&
    \frac{n_{B}^3}{3(n_{A}+n_{B})^2}.
\end{array}
\end{equation}
By direct inspection, the mole numbers given by Eqs.~(\ref{eq mole numbers1}) satisfy
Eqs.~(\ref{eq mass balance1}) and Eqs.~(\ref{eq equilibrium conditions1}) with $\Delta
g_{2,1}=0$ and $\Delta g_{1,2}=0$. Since the Gibbs free energy $G$ given by Eq.~(\ref{eq
G1}) is a strictly convex function of the mole numbers of associates, then Eq.~(\ref{eq
equilibrium conditions1}) has no other solution satisfying the mass balance.

Furthermore, if $\Delta g_{2,1}=0$ and $\Delta g_{1,2}=0$, then the molar Gibbs free
energies of associates are given by
\begin{equation}\label{eq molar g assoc}
\begin{array}{ll}
    g_{3,0}
    &=
    3g_{A},\\
    g_{2,1}
    &=
    2g_{A}+g_{B},\\
    g_{1,2}
    &=
    g_{A}+2g_{B},\\
    g_{0,3}
    &=
    3g_{B}.
\end{array}
\end{equation}
Finally, substitution of Eqs.~(\ref{eq mole numbers1}) and (\ref{eq
molar g assoc}) into Eq.~(\ref{eq G1}) yields
\begin{equation}
\label{eq G2}
    G
    =
    n_{A} g_{A}
    +
    n_{B} g_{B}
    +
    RT
    \left(
        n_{A}
        \ln
        x_{A}
        +
        n_{B}
        \ln
        x_{B}
    \right).
\end{equation}
Thus, the proposed model correctly reduces to the ideal solution
model in the case where there are no interactions between mixing
particles.

\section{Dilute solutions\label{sec dilute}}
Pelton \emph{et al.} \cite{Pelton_00} pointed out another interesting feature of the
associate model that occurs in dilute solutions. They considered the associated solution
of monoparticles $A$ and $B$ and associates $A_2B$. The highly ordered solution rich in
component $B$ consists primarily  of $B$-monoparticles and $A_2B$-associates. According
to the authors, the chemical activity $a_{B}$ of the component $B$ behaves asymptotically
as $(1-x_{A}/2)$ rather than $(1-x_{A})$ for small $x_{A}$. Thus, $\lim
\limits_{x_{A}\rightarrow 0}(da_B/dx_A)=-1/2$. Note, however, that this behaviour of
$a_{B}$ is observed only in the limiting case $\Delta g_{2,1}=-\infty$. For any finite
value, $\lim \limits_{x_{A}\rightarrow 0}(da_B/dx_A)=-1$. We present the proof of this
result for the model proposed in this paper. Using the presented technique, similar
result can be established for the example considered by Pelton \emph{et
al.}~\cite{Pelton_00}.

As obtained in Section \ref{sec description}, $a_B=x_{0,3}^{1/3}$.
The required derivative is then calculated as follows. First,
\begin{equation}\label{eq da}
    \frac{da_B}{dx_A}
    =
    \frac{1}{3}
    x_{0,3}^{-2/3}
    \frac
    {dx_{0,3}}
    {dx_A}%
    =
    \frac{1}{3}
    x_{0,3}^{-2/3}
    \left(
        \frac
        {dx_A}
        {dx_{0,3}}
    \right)^{-1}.
\end{equation}
Secondly,  assuming $\Delta g_{2,1}$ and $\Delta g_{1,2}$ finite and
denoting the right hand sides of Eqs.~(\ref{eq equlibrium
conditions1}) by
\begin{equation}
\label{epsilon}
    \epsilon_{2,1}
    \equiv
    3\exp
    \left(
        -\frac{{\Delta
        g_{2,1}}}{RT}
    \right),
    \quad
    \epsilon_{1,2}
    \equiv
    3\exp
    \left(
        -\frac{{\Delta
        g_{1,2}}}{RT}
    \right),
\end{equation}
we obtain
\begin{equation}
\label{eq xij}
\begin{array}{l}
    x_{2,1}
    =
    \epsilon_{2,1}
    {x_{3,0}^{2/3}
    x_{0,3}^{1/3}},\\
    x_{1,2}
    =
    \epsilon_{1,2}
    {x_{3,0}^{1/3}
    x_{0,3}^{2/3}}.
\end{array}
\end{equation}
Substitution of Eqs.~(\ref{eq xij}) into the mass balance
constraints of Eqs.~(\ref{eq mass balance1}), with the latter
written in terms of molar fractions, gives
\begin{equation}
\label{eq mb1}
    1
    =
    x_{3,0} + x_{0,3} + \epsilon_{2,1}{x_{3,0}^{2/3}
    x_{0,3}^{1/3}}
    +
    \epsilon_{1,2}
    {x_{3,0}^{1/3} x_{0,3}^{2/3}},
\end{equation}
\begin{equation}
\label{eq mb2}
    1-x_{A}
    =
    x_{0,3} +
    \frac{1}{3}
    \epsilon_{2,1}{x_{3,0}^{2/3} x_{0,3}^{1/3}}
+\frac{2}{3}\epsilon_{1,2}{x_{3,0}^{1/3} x_{0,3}^{2/3}}.
\end{equation}
Now, Eq.~(\ref{eq mb1}) implicitly defines $x_{0,3}$ as a function
of $x_{3,0}$, while Eq.~(\ref{eq mb2}) defines $x_{A}$ as a function
of $x_{3,0}$ and $x_{0,3}$. Differentiating Eq.~(\ref{eq mb1}) gives
\begin{equation}\label{eq mb3}
\frac{dx_{3,0}}{dx_{0,3}}=-\frac{1+\frac{2}{3}\epsilon_{1,2} t +\frac{1}{3}\epsilon_{2,1}
t^2}{1+\frac{1}{3}\epsilon_{1,2} t^{-1}+\frac{2}{3}\epsilon_{2,1} t^{-2}},
\end{equation}
where
\begin{equation}\label{eq mb3}
t \equiv \left(\frac{x_{3,0}}{x_{0,3}}\right)^{1/3}~.
\end{equation}
Differentiation of Eq.~(\ref{eq mb2}) with respect to $x_{0,3}$ and
substitution of Eq.~(\ref{eq mb3}) into the resultant expression
yields
\begin{equation}\label{eq mb4}
\begin{array}{ll}
\frac{dx_{A}}{dx_{0,3}}&=-1-\frac{4}{9}\epsilon_{1,2} t -\frac{1}{9}\epsilon_{2,1} t^2\\
&+\frac{2}{9}\frac{\left(\epsilon_{1,2} t^{-2}+\epsilon_{2,1}
t^{-1}\right)\left(1+\frac{2}{3}\epsilon_{1,2} t+ \frac{1}{3}\epsilon_{2,1} t^{2}
\right)}{\left(1+\frac{1}{3}\epsilon_{1,2} t^{-2} + \frac{2}{3}\epsilon_{2,1}
t^{-1} \right)}\\
&=-1-\frac{4}{9}\epsilon_{1,2} t -\frac{1}{9}\epsilon_{2,1} t^2\\
&+\frac{2}{9}\frac{\left(\epsilon_{1,2} +\epsilon_{2,1}
t\right)\left(1+\frac{2}{3}\epsilon_{1,2} t+
\frac{1}{3}\epsilon_{2,1} t^{2}
\right)}{\left(t^{2}+\frac{1}{3}\epsilon_{1,2} +
\frac{2}{3}\epsilon_{2,1} t \right)}.
\end{array}
\end{equation}
Note that $x_{0,3}\rightarrow 1$, $x_{3,0}\rightarrow 0$ and
$t\rightarrow 0$ as $x_{A}\rightarrow 0$. Therefore, substituting
Eq.~(\ref{eq mb4}) into Eq.~(\ref{eq da}) and taking the limit as
$x_{A}\rightarrow 0$, one verifies that $\lim
\limits_{x_{A}\rightarrow 0}(da_B/dx_A)=-1$ for any finite values of
$\Delta g_{2,1}$ and $\Delta g_{1,2}$.

Now consider the limiting case where $\Delta
g_{2,1}\rightarrow-\infty$, while $\Delta g_{1,2}$ remains finite.
Recalling Eqs.~(\ref{epsilon}) and taking the limit in Eq.~(\ref{eq
mb4}) give
\begin{equation}\label{eq mb5}
\lim \limits_{\epsilon_{2,1}\rightarrow +\infty}\frac{dx_{A}}{dx_{0,3}}=%
-\frac{2}{3}-\frac{1}{6}\epsilon_{1,2}t-\frac{1}{6}t^{3}.
\end{equation}
Finally, substituting Eq.~(\ref{eq mb5}) into Eq.~(\ref{eq da}) and
taking the limit as $x_{A}\rightarrow 0$, we obtain $\lim
\limits_{x_{A}\rightarrow 0}(da_B/dx_A)=-1/2$ as $\Delta
g_{2,1}\rightarrow-\infty$.

\newcommand{\kk}{{k_1,\ldots,k_r}}
\newcommand{\jj}{{[j]}}
\newcommand{\mm}{{(m)}}
\newcommand{\set}{{\mathfrak{S}_m}}
\newcommand{\sett}{\mathfrak{S}^\circ_m}

\section{Model equations for arbitrary number of solution components and
         arbitrary size of associates\label{sec general form}}
Consider an $r$-component solution $A_1-\ldots-A_r$ and assume that the components and
the solution itself all consist of $m$-particle associates. There are $r^m$
distinguishable types of associates. Some of these types have the same compositions. Now
consider an $m$-particle associate that consists of $k_1, \ldots, k_r$ particles of types
$A_1, \ldots, A_r$, respectively. Thus, the composition of the associate is specified by
the $r$-tuple of nonnegative integers $(k_1,\dots,k_r)$ satisfying $k_1 + \ldots + k_r =
m$. Omitting the dependence on $r$, the set of such tuplets, which represent all possible
compositions of $m$-particle associates in the $r$-component solution, is denoted by
$\set$.  Its cardinality, that is, the number ${N_\text{comp}}$ of different compositions
is computed as
\begin{equation}\label{eq N comp}
    N_\text{comp}
    =
    \frac
    {(m+r-1)!}
    {m!(r-1)!}~.
\end{equation}
Consider a pure solution component $A_i$. Its $m$-particle
associates all have the composition
\begin{equation}
    \sigma_i
    \equiv
    (
        \small{\underset{i-1}
        {\underbrace{0,\dots,0}}},
        m,
        \small
        {\underset{r-i}
        {\underbrace{0,\dots,0}}}).
\end{equation}
The compositions $\sigma_1, \ldots, \sigma_r$ and corresponding
associate types are referred to as \emph{pure}. The complementary
set $\set\setminus \{\sigma_1, \ldots, \sigma_r\}$ of \emph{mixed}
compositions, containing two or more different particle species,  is
written briefly as $\sett$. Thus, $\sett$ is constituted by those
$r$-tuples $(k_1, \ldots, k_r)$ from $\set$ with at least two
nonzero entries.

In one mole of the solution component $A_i$, there are $1/m$ moles
of $m$-particle associates of composition $\sigma_i$. Thus, their
molar Gibbs free energy $g_{\sigma_i}$ is
\begin{equation}
\label{eq g comp}
    g_{\sigma_i}
    =
    m g_i,
\end{equation}
where $g_i$ is the molar Gibbs free energy of the solution component
$A_i$.

In general, the total number $N_\kk$ of distinguishable types of
associates which have the same composition $\kk$ is described by the
multinomial coefficient
\begin{equation}\label{eq N_kk_fixed_composition}
    N_\kk
    =
    \frac
    {m!}
    {\prod\limits_{i=1}^{r}
    k_i!}.
\end{equation}
We assume that all distinguishable types of the associates of
composition $(\kk)$ are endowed with $J_\kk$ different values of the
molar Gibbs free energy. Let $d_\kk^\jj$ be the number of the
associate types at the $j$th energy level $g_\kk^\jj$, that is, the
multiplicity of the level, so that
\begin{equation}
    \sum\limits_{j=1}^{J_\kk}
    d_\kk^\jj
    =
    N_\kk.
\end{equation}
All the $d_\kk^\jj$ associate types have the same molar Gibbs free
energy of formation $\Delta g_\kk^\jj$ according to the reaction
\begin{equation}\label{eq reaction general}
    \frac
    {k_1}
    {m}
    (A_1)_m
    +
    \ldots
    +
    \frac{k_r}{m}
    (A_r)_m
    \rightleftarrows
    \left(
        (A_1)_{k_1}
        \ldots
        (A_r)_{k_r}
    \right)^\jj.
\end{equation}
More precisely, the Gibbs free energy of the reaction is defined by
\begin{equation}
\label{eq Delta g general}
    \Delta g_\kk^\jj
    =
    g_\kk^\jj
    -
    \frac{k_1}{m}
    g_{\sigma_1}
    -
    \ldots
    -
    \frac{k_r}{m}
    g_{\sigma_r}.
\end{equation}
The molar Gibbs free energies $g_\kk^\jj$ of mixed associates of
composition $(\kk) \in \sett$, or alternatively, the corresponding
formation energies $\Delta g_\kk^\jj$ from Eq.~(\ref{eq Delta g
general}) are adjustable parameters of the model.

Now let $n_\kk^\jj$ denote the mole number of the associates of
composition $(\kk)$ at the $j$th energy level. The mass balance
constraints then read
\begin{equation}
\label{eq mass balance general}
    n_i
    =
    \sum\limits_{(\kk) \in \set}
    k_i
    \sum\limits_{j=1}^{J_\kk}
    n_\kk^\jj,
    \qquad
    i = 1, \ldots, r,
\end{equation}
where $n_i$ is the mole number of $A_i$ particles in the solution.
One verifies that
\begin{equation}\label{eq nt}
    \sum\limits_{i=1}^{r}
    n_i
    =
    m
    \sum\limits_{(\kk) \in \set}
    \sum\limits_{j=1}^{J_\kk}
    n_\kk^\jj
    \equiv
    m \,n_\text{tot}.
\end{equation}
Note that the total number of associates $n_\text{tot}$ is independent of composition of
the solution since they are assumed to be of equal size. The Gibbs free energy of the
solution is computed as
\begin{equation}\label{eq G general}
\begin{array}{lcl}
    G
&= &
    \sum\limits_{(\kk) \in\set}%
    \sum\limits_{j=1}^{J_\kk}
    n_\kk^\jj
    g_\kk^\jj\\
&+ &
    RT
    \sum\limits_{(\kk) \in\set}
    \sum\limits_{j=1}^{J_\kk}
    n_\kk^\jj
    \ln
    \left(
        \frac
        {x_\kk^\jj}
        {d_\kk^\jj}
    \right),
\end{array}
\end{equation}
where
\begin{equation}
    x_\kk^\jj
    =
    \frac
    {n_\kk^\jj}
    {n_\text{tot}}
\end{equation}
is the molar fraction of an appropriate associate type.

Eqs.~(\ref{eq mass balance general}) define $n_{\sigma_i}$ as a linear function of the
mole number $n_i$ of the solution component $A_i$  and of the mole numbers of the mixed
associates $n_\kk^\jj$, where $(\kk) \in \sett$. The equilibrium values  of $n_\kk^\jj$,
which are internal variables of the model, are determined by minimising the Gibbs free
energy of the solution at constant $n_{1}, \ldots, n_{r}$, subject to the mass balance
constraints of Eqs.~(\ref{eq mass balance general}). The minimum is found by setting
\begin{equation}\label{eg dG dn general}
    \left(
        \frac
        {\partial G}
        {\partial n_\kk^\jj}
    \right)_{n_1,\ldots,n_r}
    =
    0
\end{equation}
for all $(\kk) \in \sett$ and for all $\jj=1,\dots,J_\kk$. Recalling that $n_{\sigma_i}$
are functions of $n_\kk^\jj$, where $(\kk) \in \sett$, and using the chain rule,
\begin{equation}
\label{eq dG dn1 general}
    \frac
    {\partial G}
    {\partial n_\kk^\jj}
    +
    \sum\limits_{i=1}^r
    \frac
    {\partial G}
    {\partial n_{\sigma_i}}
    \frac
    {\partial n_{\sigma_i}}
    {\partial n_\kk^\jj}
    =
    0.
\end{equation}
Substitution of Eqs.~(\ref{eq mass balance general}) and (\ref{eq G general}) into
Eq.~(\ref{eq dG dn1 general}) gives
\begin{equation}\label{eq equlibrium conditions general}
    \frac
    {n_\kk^\jj}
    {\prod\limits_{i=1}^{r} (n_{\sigma_i})^{k_i/m}}
    =
    d_\kk^\jj
    \exp
    \left(
        -\frac
        {\Delta g_\kk^\jj}
        {RT}
    \right).
\end{equation}
Using Eqs.~(\ref{eq mass balance general}), one verifies that
$n_{\sigma_i}$ is a function of $n_i$ and of $n_\kk^\jj$, where
$(\kk) \in \sett$. The chemical potential $\mu_1$ of the solution
component $A_1$ can now be calculated as follows:
\begin{equation}
\label{eq mu general} \!\!
\begin{array}{lcl}
    \mu_1
    & = &
    \left(
        \frac
        {\partial G}
        {\partial n_{1}}
    \right)_{n_2,\dots,n_r}\\
    & = &
    \sum\limits_{(\kk) \in \sett}
    \sum\limits_{j=1}^{J_\kk}
    \frac
    {\partial G}
    {\partial n_\kk^\jj}
    \frac
    {\partial n_\kk^\jj}
    {\partial n_1}\\
    & + &
    \sum\limits_{i=1}^{r}
        \frac
        {\partial G}
        {\partial n_{\sigma_i}}
        \left(
            \frac
            {\partial n_{\sigma_i}}
            {\partial n_1}
            +
            \sum\limits_{(\kk) \in \sett}
            \sum\limits_{j=1}^{J_\kk}
            \frac
            {\partial n_{\sigma_i}}
            {\partial n_\kk^\jj}
            \frac{\partial n_\kk^\jj}
            {\partial n_1}
        \right)\\
    & = &
    \frac
    {\partial G}
    {\partial n_{\sigma_1}}
    \frac
    {\partial n_{\sigma_1}}
    {\partial n_1}
    +
    \sum\limits_{(\kk) \in \sett}
    \sum\limits_{j=1}^{J_\kk}
    \left(
        \frac
        {\partial G}
        {\partial n_\kk^\jj}
    \right.\\
    &+&
    \left.
        \sum\limits_{i=1}^r
        \frac
        {\partial G}
        {\partial n_{\sigma_i}}
        \frac
        {\partial n_{\sigma_i}}
        {\partial n_\kk^\jj}
    \right)
        \frac{\partial n_\kk^\jj}
        {\partial n_1}
    \\
    & = &
    \frac
    {\partial G}
    {\partial n_{\sigma_1}}
    \frac
    {\partial n_{\sigma_1}}
    {\partial n_1}
\end{array}
\end{equation}
Here, Eq.~(\ref{eq dG dn1 general}) have been used. Using Eqs.~(\ref{eq g comp}) and
(\ref{eq mass balance general}), one verifies that
\begin{equation}
    \mu_1
    =
    g_1
    +
    \frac{1}{m}
    RT
    \ln
    x_{\sigma_1}.
\end{equation}
The chemical potentials of the other components are calculated in a
similar way.

Now assume that all the associate types of composition $(\kk)$ have the same Gibbs free
energy of formation for any $(\kk) \in \set$, so that $J_\kk =1$ and
$d_\kk^1=m!/\prod\limits_{i=1}^{r}k_i!$. This assumption, which can be subsequently
refined in thermodynamic model optimisation of real chemical systems, allows
substantially reduce the number of adjustable parameters of the model. Since the Gibbs
free energies of pure associates are fixed, the number of adjustable parameters is equal
to $(N_\text{comp}-r)$, where $N_\text{comp}$ is given by Eq.~(\ref{eq N comp}). Then,
omitting the superscript $\jj$, Eq.~(\ref{eq G general}) reduces to
\begin{equation}
\label{eq G1 general}
\begin{array}{ll}
    G
    &=
    \sum\limits_{(\kk) \in\set}
    n_\kk
    \left(
        g_\kk
        +
        RT
        \ln
        \left(
            \frac
            {x_\kk {\prod\limits_{i=1}^{r}k_i!}}{m!}
        \right)
    \right)
\end{array}
\end{equation}
Similarly to Section \ref{sec ideal case}, consider the case when
all Gibbs free energies $\Delta g_\kk$ of formation of associates
are equal to zero, so that  particles of the solution components are
mixed ideally. In this case, the composition of $m$ randomly
selected particles follows the multinomial distribution; see Ref.
\cite{Feller_1} for more details. The molar fraction of the
associates with composition $(\kk) \in \set$ is expressed as
\begin{equation}\label{eq probabilities general}
    x_\kk
    =
    m!
    \prod\limits_{i=1}^{r}
    \frac
    {x_i^{k_i}}
    {k_i!}
\end{equation}
and hence, their mole number is
\begin{equation}\label{eq n_kk}
    n_\kk
    =
    \frac
    {(m-1)!}
    {\left(
        \sum\limits_{i=1}^{r}
        n_i
        \right)^{m-1}
    }
    \prod\limits_{i=1}^{r}
    \frac
    {n_i^{k_i}}
    {k_i!}.
\end{equation}
Here, $n_i$ and $x_i$ are the mole number and the molar fraction of the solution
component $A_i$. One verifies that the mole numbers given by Eq.~(\ref{eq n_kk}) describe
the solution of Eq.~(\ref{eq equlibrium conditions general}) subject to the mass balance
constraints of Eqs.~(\ref{eq mass balance general})  in the case $\Delta g_\kk =0$. The
uniqueness of the solution is ensured by the strict convexity of the Gibbs free energy
$G$ given by Eq.~(\ref{eq G1 general}) in the variables $n_\kk$.

From Eqs.~(\ref{eq g comp}) and (\ref{eq Delta g general}), the molar Gibbs free energy
of the associates of the composition $(\kk)$ is given by
\begin{equation}\label{eq g_kk}
g_\kk = \sum\limits_{i=1}^{r}{k_i}g_i~,
\end{equation}
where, $g_i$ is the molar Gibbs free energy of the solution
component $A_i$. Substitution of Eqs.~(\ref{eq probabilities
general})-(\ref{eq g_kk}) into Eq.~(\ref{eq G1 general}) and a
straightforward, though lengthy, verification shows that the
proposed model correctly reduces to the $r$-component ideal solution
model in this case. That is,
\begin{equation}\label{eq G2 general}
    G
    =
    \sum\limits_{i=1}^{r}
    n_{i}
    \left(
        g_{i}
        +
        RT
        \ln
        x_{i}
    \right).
\end{equation}

From this reduction, it is immediately follows that the model with associates of size $r$
can be reproduced by the models with associates of size $2r$, $3r$ and so on. For
example, consider the model with associates of size $2r$. Any $2r$-associate is a
combination of two $r$-associate. Now assume that there is no Gibbs free energy change on
forming any $2r$-associate from corresponding two $r$-associates. As demonstrated above,
the model with $2r$-associate reduces to the ideal mixture of $r$-associates. Thus, the
model with $2r$-associates is more general and include the model with $r$-associates as a
particular case.

\section{Effective adjustable parameters of the model}
The adjustable parameters of the model related to associates of the composition $(\kk)$
are $\Delta g_\kk^\jj$, where $\jj=1,\dots,J_\kk$. One should also take into account the
multiplicity of the energy levels $d_\kk^\jj$. In general, the number of adjustable
parameters of the model increases exponentially with the increase in size of associates
$m$. However, the number of adjustable parameters can be substantially reduced without
loss of generality of the model as described below.

Consider associates with the composition $(\kk)$. In general, $J_\kk$ energy levels are
possible for these associates. The mole number $n_\kk$ of \emph{all} associates with the
composition $(\kk)$ is given by
\begin{equation}\label{e1}
    n_\kk
    =
    \sum\limits_{j=1}^{J_\kk}
        n_\kk^\jj.
\end{equation}
The mass balance constraints of Eqs.~(\ref{eq mass balance general}) take the form
\begin{equation}
\label{eq mass balance general 1}
    n_i
    =
    \sum\limits_{(\kk) \in \set}
    k_i
    n_\kk,
    \qquad
    i = 1, \ldots, r.
\end{equation}
Summing up over $\jj$ in Eq.~(\ref{eq equlibrium conditions general}), one verifies that
the equilibrium value of $n_\kk$ is calculated as
\begin{equation}\label{e2}
    \frac
    {n_\kk}
    {\prod\limits_{i=1}^{r} (n_{\sigma_i})^{k_i/m}}
    =
    Z_\kk
\end{equation}
where
\begin{equation}\label{Z def}
    Z_\kk
    \equiv
    \sum\limits_{j=1}^{J_\kk}
        d_\kk^\jj
        \exp
        \left(
            -\frac
            {\Delta g_\kk^\jj}
            {RT}
        \right).
\end{equation}

The function $Z_\kk$ defined by Eq.~(\ref{Z def}) plays a role of the partition function
which describes the distribution of associates of the composition $(\kk)$ over energy
levels. That is,
\begin{equation}\label{e4}
\frac{n_\kk^\jj}
     {n_\kk}%
=%
\frac{x_\kk^\jj}
     {x_\kk}
=%
\frac{d_\kk^\jj}
     {Z_\kk}
\exp
    \left(
        -\frac{\Delta g_\kk^\jj}{RT}
    \right)
\end{equation}
In fact, $Z_\kk$ is a single effective adjustable parameter related to associates of the
composition $(\kk)$. The other thermodynamic parameters of the model can be expressed in
terms of $Z_\kk$, where $\kk \in \set$, and their derivatives. Indeed, using
Eqs.~(\ref{eq g comp}), (\ref{eq Delta g general}), (\ref{eq mass balance general 1}) and
(\ref{e4}), one verifies that the Gibbs free energy of the solution given by Eq.(\ref{eq
G general}) takes the form
\begin{equation}\label{e5}
    G
    =
    \sum\limits_{i=1}^{r}
        n_i g_i
    +
    RT
    \sum\limits_{(\kk)\in \set}
       n_\kk \ln
       \left(
            \frac{x_\kk}{Z_\kk}
       \right).
\end{equation}

According to Eq.~(\ref{Z def}), ${Z_\kk}$ varies from zero to infinity. When the
associates of the composition $(\kk)$ are prohibited energetically, that is $(\Delta
g_\kk^\jj \rightarrow + \infty)$ for all $\jj$, ${Z_\kk}$ approaches zero. If the
associates of the composition $(\kk)$ is highly preferable, ${Z_\kk}$ approaches
infinity. Similar to the Gibbs free energies of associates, the optimal values of the
effective adjustable parameters ${Z_\kk}$ can be determined by the trial-and-error
procedure which is conventionally used in thermodynamic model optimisation of real
chemical systems.

\section{Excess Gibbs energy terms}
Similarly to the modified associate species model \cite{Besmann06}, regular or, in
general, polynomial, excess Gibbs free energy terms can be included into the proposed
formalism. These terms take into account interactions between associates. A probabilistic
interpretation of the polynomial excess Gibbs free energy terms is presented in Ref.
\cite{Saulov_multicomponent}. In fact, this interpretation provides a theoretical
justification for such terms. Treating the associates as particles, the results of
Ref.~\cite{Saulov_multicomponent} are applicable to the model presented in this paper.

Note that it is desirable to use the excess terms only for ``fine
tuning'' of the model, while the main adjustable parameters are the
molar Gibbs free energies of associates. The following two
conditions on the absolute value of the interaction parameters
should be satisfied; see, for example, the monograph by Prigogine
and Defay \cite{PrigogineDefay} for more details.
\begin{enumerate}
\item
The absolute values of the interaction parameters should be small
compared with the molar Gibbs free energies of the associates. As
pointed out by Prigogine and Defay \cite{PrigogineDefay}, if the
interaction between associates, say ``C1'' and ``C2'', is
sufficiently strong to alter the vibrational and rotational states
of the associates, then the associate ``C1C2'' is included into the
set of associates by the definition of the classical associate
model. In the framework of the proposed formalism, one should
consider the associates of larger size.

\item
The values should also be small in comparison with $RT$. Otherwise, the assumption of
ideal mixing of associates is less justified. Again, larger associates should be taken
into account.
\end{enumerate}

Note, however, that the second condition is sometimes relaxed in
order to fit the experimental data available for real solutions.
This is the case, for example, for the solutions with immiscibility,
which is the result of relatively weak, compared with the Gibbs free
energies of associates, repulsive interactions between the
associates.

\section{Discussion on applicability of the model}
The suggested modified associate formalism belongs to associate-type models. As a result,
the range of applicability of the formalism is at least the same as that of the classical
associate model or the associate species model. There are, however, some distinctions in
using the suggested formalism and the previous modifications of the associate model.
These distinctions are discussed below. Since the present study is intended as a
theoretical introduction to  the modified associate formalism, the results on its
application to real chemical systems will be reported elsewhere.

In contrast to the previous modifications, where associates of arbitrary compositions can
be included into the set of associates, the compositions of associates are defined by
their size in the framework of the modified associate formalism. Therefore, a modeller
should pay special attention to selection of the size of associates. If experimental
information about compositions of associates that present in the solution phase is
available, this information should clearly be taken into account. The size of associates
should be large enough to incorporate those compositions.

When the suggested formalism is applied for thermodynamic description of multicomponent
solution phases, the size of associate should be large enough to incorporate the
associates of the compositions, which coincide with those of maximum ordering in all
binding binary systems. For example, consider the ternary system $A-B-C$ and assume that
in the binary system $B-C$ the composition of maximum ordering is that of the associate
$B_2C$, while in the systems $A-B$ and $A-C$ maximum ordering occurs at the compositions
of the associates $AB$ and $AC$, respectively. To incorporate the required composition,
the size of associates should be divisible by 2 and by 3. Therefore, the size of
associates for the ternary system $A-B-C$  should be at least 6.

It is desirable to use 6-particle associates for thermodynamic model optimisation for all
the binary systems. There is, however, no need to use 6-particle associates from the
beginning. Let us assume that the systems A-B and A-C are initially optimised with
2-particle associate, while the system B-C is optimised with 3-particle associate. As
discussed in the end of Section 7, the model with associates of size $r$ can be
reproduced by the models with associates of size $2r$, $3r$ and so on. Using this
property of the proposed formalism, the descriptions of the binary systems with 2- and
3-particle associates can be replaced by the equivalent descriptions with 6-particle
associates in a straightforward way. Then, the binary systems can be combined into the
ternary one.

As a demonstrational example, consider the system $A-B$. When the system is optimised
with 2-particle associates the Gibbs free energies of the associates $g_{[AA]}$,
$g_{[BB]}$, $g_{[AB]}$ and $g_{[BA]}$ are known. Due to spatial symmetry, $g_{[AB]} =
g_{[BA]}$. In the equivalent description of the system $A-B$, 6-particle associates are
formed from three 2-particle associates with no Gibbs free energy changes on such
formations. The associate of the composition $A_6$ is formed from three $[AA]$-associates
with $g_{A_6}=3g_{[AA]}$. The associate of the composition $A_5B$ is formed ether from
two $[AA]$-associates and $[AB]$-associates or from two $[AA]$-associates and
$[BA]$-associates. Then, recalling that $g_{[AB]} = g_{[BA]}$, $A_5B$-associates have
only one energy level $g_{A_5B}=2g_{[AA]}+g_{[AB]}$ with multiplicity 6. The
$A_4B_2$-associate can be formed from two $[AA]$-associates and one $[BB]$-associate with
the energy level $g^{[1]}_{A_4B_2}=2g_{[AA]}+g_{[BB]}$ and the multiplicity
$d^{[1]}_{A_4B_2}=3$. Alternatively, $A_4B_2$-associate can be form from one
$[AA]$-associate and either two $[AB]$-associates or two $[BA]$-associates or one
$[AB]$-associate and one $[BA]$-associate. In this case, the energy level is
$g^{[2]}_{A_4B_2}=g_{[AA]}+2g_{[AB]}$, and its multiplicity is $d^{[2]}_{A_4B_2}=12$. In
a similar way, one verifies that $A_3B_3$-associates have two energy levels:
$g^{[1]}_{A_3B_3}=g_{[AA]}+g_{[BB]}+g_{[AB]}$ with the multiplicity $d^{[1]}_{A_3B_3}=12$
and $g^{[2]}_{A_3B_3}=3g_{[AB]}$ with the multiplicity $d^{[2]}_{A_3B_3}=8$.
$A_2B_4$-associate also have two energy levels: $g^{[1]}_{A_2B_4}=g_{[AA]}+2g_{[BB]}$
with the multiplicity $d^{[1]}_{A_2B_4}=3$ and $g^{[2]}_{A_2B_4}=g_{[BB]}+2g_{[AB]}$ with
the multiplicity $d^{[2]}_{A_2B_4}=12$, while $AB_5$-associates have only one energy
level $g_{AB_5}=2g_{[BB]}+g_{[AB]}$ with the multiplicity equals to 6. $B_6$-associates
also have only one energy level $g_{B_6}=3g_{[BB]}$ with the multiplicity equals to 1.
The binary systems $A-C$ and $B-C$ are treated similarly.

Previous modifications of the associate model allow more flexibility in fitting
experimental data compared with the suggested approach, since stoichiometry of associates
can be arbitrarily altered to improve the fit. This additional flexibility can be helpful
for modelling a binary system with high degree of ordering, where the described paradox
is of relatively low practical importance. However, when such a binary system is combined
with other binaries to form multicomponent model, arbitrary selection of associate
stoichiometry can result in the negative consequences of the entropy paradox.

Similar to the previous modifications, a miscibility gap can not be reproduced without
excess Gibbs free energy terms in the proposed formalism. However, as demonstrated by
Besmann et el.~\cite{Besmann06} for the associate species model, immiscibility, which is
the result of repulsive interactions between associate (or associate species), can be
accurately represented by the associate-type models with the polynomial excess Gibbs free
energy. Treating associates (or associate species) as particles and using the results of
Ref.~\cite{Saulov_multicomponent}, one verifies that the coefficients of the polynomial
excess Gibbs free energy explicitly relate to energies of interactions between
associates. It is important to note that, in the case of the modified quasichemical
model~\cite{Pelton_00}, immiscibility is described by empirical polynomial expansions of
the Gibbs free energy of the quasichemical reaction. In contrast to associate-type
models, physical meaning of the coefficients of such expansions is not clear.

Another distinction of the proposed formalism from the previous modifications is related
to multicomponent associates, that is, the associates which consist of particles of three
or more types. In the previous modifications, no multicomponent associates are initially
considered. This implies that all multicomponent associates are assumed to have infinite
positive Gibbs free energy. Multicomponent associates are usually included into the
model, when such associates are required to fit available experimental data in
multicomponent systems. In contrast, the compositions of multicomponent associates are
prescribed by the selected size of associate in the proposed formalism. It would be
beneficial to develop a method for estimating the Gibbs free energies of the
multicomponent associates from those of binary associates. The development of such a
method, however, is the topic for separate study. As an initial approximation, the Gibbs
free energies of formation of the multicomponent associates from pure associates can be
set to zero. If experimental data are available in multicomponent systems, the Gibbs free
energies of the multicomponent associates can be adjusted in a conventional
trial-and-error procedure.

A possibility to select compositions of multicomponent associates arbitrarily, which
exists in the previous modifications, could provide more flexibility in data fitting
compared with the fixed set of compositions. It could be the case, that arbitrarily
selected additional multicomponent associates provide better fitting to particular
experimental data. One should recognise, however, that the price for the possibility to
select the compositions of associates arbitrarily is the entropy paradox, which
undermines the fundamental predictive capabilities of the model. In our opinion, the
necessity of additional multicomponent associates for fitting the available experimental
data indicates that the size of associates should be increased to incorporate the
required compositions.

\section{Conclusion}
In this paper, the Modified Associate Formalism has been proposed for thermodynamic
modelling of solutions. The presented approach is free from the entropy paradox and
correctly reduces to the ideal solution model, where there are no Gibbs free energy
changes on forming associates of different types.

Asymptotic behaviour of chemical activities of solution components
in binary dilute solutions has been investigated. It has been
demonstrated that the derivative of the chemical activity of a
solution component in its molar fraction at terminal composition has
the expected value for any finite value of the  Gibbs free energies
of formation of associates.

\section*{Acknowledgment}
The present work has been supported by the Australian Research Council.

\bibliographystyle{elsart-num}


\end{document}